# Thermal effect on magnetic parameters of high-coercivity cobalt ferrite

Chagas, E. F.[1], Ponce, A. S.[1], Prado, R. J.[1], Silva, G. M.[1], J. Bettini[3] and Baggio-Saitovitch, E.[2]

[1]*Instituto de Física, Universidade Federal de Mato Grosso, 78060-900, Cuiabá-MT, Brazil*
[2]*Centro Brasileiro de Pesquisas Físicas, Rua Xavier Sigaud 150 Urca. Rio de Janeiro, Brazil.*
[3]*Laboratório Nacional de Nanotecnologia, Centro Nacional de Pesquisa em Energia e Materiais, 13083-970, Campinas, Brazil*

Phone number: 55 65 3615 8747

Fax: 55 65 3615 8730

Email address: efchagas@fisica.ufmt.br

**Abstract**

We prepared very high-coercivity cobalt ferrite nanoparticles using short-time high-energy mechanical milling. After the milling process, the coercivity of the nanoparticles reached 3.75 kOe - a value almost five times higher than that obtained for the non-milled sample. We performed a thermal treatment on the milled sample at 300, 400, and 600 $^o$C for 30 and 180 mins, and studied the changes in the magnetic parameters due to the thermal treatment using the hysteresis curves, Williamson-Hall analysis, and transmission electron microscopy. The thermal treatment at 600 $^o$C causes a decrease in the microstructural strain and density of structural defects resulting in a significant decrease in coercivity. Furthermore, this thermal treatment increases the size of the nanoparticles and, as a consequence, there is a substantial increase in the saturation magnetization. The coercivity and the saturation magnetization are less affected by the thermal treatment at 300 and 400 $^o$C.

# Introduction

The hard magnetic compound $CoFe_2O_4$ (cobalt ferrite) presents interesting characteristics, such as chemical stability, electrical insulation, high magnet-elastic effect [1], thermal chemical reduction [2-4], moderate saturation magnetization ($M_S$) and hard coercivity ($H_C$). Due to these characteristics, the cobalt ferrite is a promising material for several technological applications, such as high density magnetic storage [5], electronic devices, biomedical applications [6,7], and permanent magnets [8].

To permanent magnets applications two quantities are fundamental: coercivity and remanence ($M_R$). Both compose the figure of merit in a hard magnetic material, the quantity energy product $(BH)_{max}$, that gives an idea of the amount of energy that can be stored in the material. Several works report tuning of the coercivity using different methods: thermal annealing [9], capping [10] and mechanical milling treatment [11,12] of the grains. Liu et al. increased the $H_C$ of cobalt ferrite, from 1.23 to 5.1 kOe, with a relatively small decrease in $M_S$ due the decrease of the grain size. They used a brief (1.5h) mechanical milling process on relatively large particles (average grain size of 240nm). In a previous work [12], using mechanical milling, we reached an increase of up to 4.2 kOe in the coercivity of cobalt ferrite nanoparticles with an average grain size of about 23 nm.

To increase the saturation magnetization and remanence is more complicated, but the exchange spring effect can be used to increase the saturation magnetization of some nanomaterials [4,13-15]. The exchange spring effect was observed for the first time in 1989 by Coehoorn et al. [16], and explained in 1991 by E. F. Kneller and R. Hawig [13], whor argue that, under certain conditions, hard and soft magnetic materials may present exchange coupling when in nanocomposite form. In this case, the high anisotropy of the hard material gives to nanocomposite a high coercivity, and high saturation magnetization of the soft material gives the high $M_S$, substantially increasing the product $(BH)_{max}$ when compared with any one of the individual phases of the nanocomposite.

Cobalt ferrite is a promising material for obtaining optimized exchange-spring magnets [4,15] due its characteristic of thermal chemical reduction, cited above. This property allows the transformation of $CoFe_2O_4$ into $CoFe_2$ (a soft ferromagnetic material with high $M_S$ value of about 230 emu/g [17]) at moderate/high temperatures, and controlling the ratio of $CoFe_2$ (in the nacomposite $CoFe_2O_4/CoFe_2$) must be optimizing the quantity $(BH)_{max}$ [15]. All processes utilized to obtain the nanocomposite $CoFe_2O_4/CoFe_2$ employ thermal treatment [4,15]. Is reasonable to suppose that more hard is the precursor material ($CoFe_2O_4$) better the nanocomposite will be for permanent magnet application.

To investigate the thermal effect on magnetic parameters, we prepared high-coercivity cobalt ferrite using mechanical milling and annealing it at various temperatures for different periods of time.

## Experimental Procedure

The cobalt ferrite nanopowder (sample CF0) was synthesized by a pH-controlled nitrate-glycine gel-combustion process [18,19]. High-purity (99.9%) raw compounds were used. The synthesis process was adjusted to obtain 3 g of the final product.

Cobalt nitrate and iron nitrate (VETEC, Brazil) were dissolved in 450 ml of distilled water in a ratio corresponding to the selected final composition. Glycine (VETEC, Brazil) was added in a proportion of one and half moles per mole of metal atoms, and the pH of the solution was adjusted with ammonium hydroxide (25%, Merck, Germany) in the range of 3 to 7. The pH was tuned as closely as possible to 7, taking care to avoid precipitation. The resulting solution was concentrated by evaporation using a hot plate at 300°C until a viscous gel was obtained. This hot gel finally burnt out as a result of a vigorous exothermic reaction. The system remained homogeneous throughout the entire process and no precipitation was observed. Finally, the as-reacted material was calcined in air at 700°C for 2 h in order to remove the organic residues.

In this work, CF0 is the sample as synthesized and CF is the same sample after milling, but without any thermal treatment. The sample CF300_180 was thermally treated at 300 C (first number) for 180 minutes (second number). The naming of the other samples followed the same labeling rule.
The mechanical processing used to increase the coercivity of the samples was described in detail in a previous work [12]. A Spex 8000 high-energy mechanical ball miller, equipped with 6 mm diameter zirconia balls, was used for 1.5 h for all samples with ball/sample mass ratio of about 1/7.

The morphology and particle size distribution of the samples were examined by direct observation via transmission electron microscopy (TEM) using a JEOL-2100 apparatus installed at LNNano / LNLS – Campinas – Brazil, working at 200 kV.

The crystalline phases of the cobalt ferrite nanoparticles were identified by X-ray powder diffraction (XPD) patterns, obtained on a Shimadzu XRD-6000 diffractometer installed at the "*Laboratório Multiusuário de Técnicas Analíticas*" (LAMUTA / UFMT – Cuiabá - MT – Brazil). It was equipped with graphite monochromator and conventional Cu tube (0.154178 nm), working at 1.2 kW (40 kV, 30 mA). Bragg-Brentano geometry was used. For the Williamson-Hall analysis [20-22], the instrumental broadening of the apparatus was determined using a $Y_2O_3$ diffraction pattern as standard.

Magnetic measurements were carried out using a vibrating sample magnetometer (VSM) model VersaLab Quantum Design, installed at CBPF, Rio de Janeiro-RJ – Brazil. Experiments were done at room temperature and using a magnetic field up to 2.7 T.

## Results and discussions

The XPD patterns obtained from samples CF, CF600_180, CF600_30, CF400_30, CF300_180, and CF300_30 (see figure 1) confirm that all samples are $CoFe_2O_4$ with the expected inverse spinel structure. Furthermore, these measurements indicate the absence of any other phases or contamination after milling and thermal treatments.

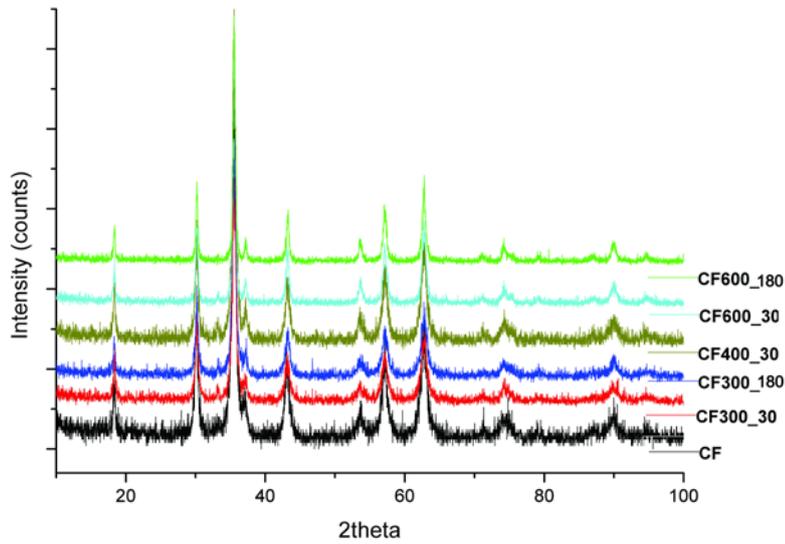

Figure 1 - XPD patterns of the samples CF, CF600_180, CF600_30, CF400_30, CF300_180, and CF300_30.

The magnetic behaviour was evaluated via measurements of the hysteresis loops obtained by VSM at room temperature. A substantial increase in coercivity was obtained for the milled sample CF ($H_C$=3.75kOe) compared with the non-milled sample (0.76kOe). This $H_C$ value is close to that obtained in a previous work (4.2 kOe) [12].

To facilitate the analysis, we summarized in Table 1 the results of $H_C$, $M_S$, and $M_R$ for all samples. Analysing these three parameters, one can note that there are two clear tendencies: first is the decrease of the $H_C$, and second is the increase of the $M_S$ and $M_R$ with the increase in temperature and time of the thermal treatment.

Table 1 - Magnetic parameters

| Sample | $H_C$ (kOe) | $M_S$* (emu/g) | $M_R$ (emu/g) |
|---|---|---|---|
| CF0 | 0.76 | - | - |
| CF | 3.75 | 57 | 33.0 |
| CF300_30 | 3.46 | 57 | 32.8 |
| CF300_180 | 3.39 | 58 | 34.0 |
| CF400_30 | 3.00 | 62 | 36.7 |
| CF600_30 | 2.24 | 66 | 39.6 |
| CF600_180 | 1.93 | 70 | 42.0 |

*value of magnetization obtained at magnetic field equal to 2.7 T.

For clarity, we show in Figure 2 the hysteresis curve for only three samples: CF, CF300_30, and CF600_180. Despite the tendency for the $H_C$ to decrease in the samples treated at 300°C, the decrease was only about 8 % and 10% in samples treated for 30 min and 180 min respectively, and no significant change in $M_S$ and $M_R$ was observed. However, in the sample treated at 600 °C for 180 min, $H_C$ decreased almost 50%, and saw a significant and simultaneous increase in $M_S$. The samples CF600_30 and CF400_30 also followed the tendencies of decreasing $H_C$ and increasing $M_S$ when increasing the temperature of the thermal treatment.

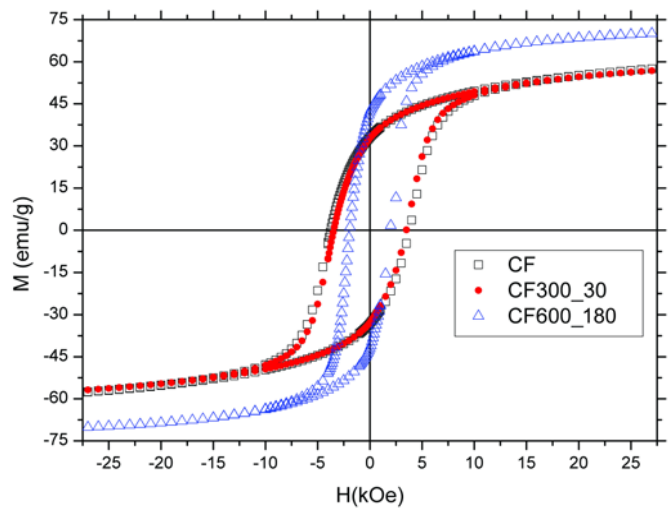

Figure 2 – Hysteresis loop of cobalt ferrite nanocomposites CF, CF300_30 and CF600_180.

To explain these results it is necessary first to understand the cause of the increase in coercivity in the milled cobalt ferrite. This increase was studied in a previous work [12] and by Liu et al. [11], and two possible factors were identified: the increase of the stress anisotropy (due the microstructural strain); and the increase of the density of structural defects that causes the increase of the centres of domains wall pinning. Therefore, we associate the decrease in coercivity following the thermal treatment observed in this work with decreases in the stress anisotropy and density of structural defects.

The thermal treatment causes the increase of the nanoparticle grains and consequently decreases the ratio between surface and bulk atoms in the material. The surface magnetic atoms, due the magnetic disorder caused by finite-size behaviour, contribute a smaller portion to the macroscopic magnetic moment when compared to the bulk atoms [23,24]. We associate this effect with the increase in the $M_S$ observed to samples treated at 400 and 600$^o$C. The same effect, but in an opposite direction (i.e., a decrease in the $M_S$ caused by the decrease of the nanoparticle grains) was observed by Liu and Ding [11] to milled cobalt ferrite.

To confirm our explanations of the decrease in $H_C$ and the increase in $M_S$ when increasing the temperatures of thermal treatment, a Williamson-Hall analysis of the XPD data and TEM measurements were performed. The Williamson-Hall analysis in Figure 3 shows that the broadening of the diffraction peaks after milling evidences both grain size reduction and an increase in microstructural strain.

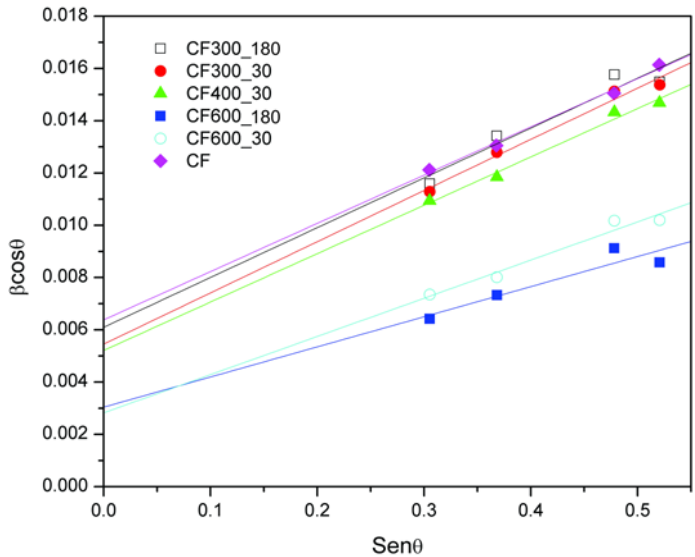

Figure 3 – Williamson-Hall analysis of cobalt ferrite nanocomposite milled and annealed at various temperatures for different periods of time.

To facilitate the analysis of the influence of structural parameters on the magnetic behaviour of the samples, we included in Table III results obtained from the Williamson-Hall analysis together with coercivity and saturation magnetization.

Table 2 - Results obtained from Williamson-Hall analysis together with coercivity and saturation magnetization.

| Sample | Average crystallite size (nm) | $M_S$ (emu/g) | Strain (%) | $H_C$ (kOe) |
|---|---|---|---|---|
| CF | 24 | 57 | 0.92 | 3.75 |
| CF300_30 | 28 | 57 | 0.98 | 3.46 |
| CF300_180 | 25 | 58 | 0.95 | 3.39 |
| CF400_30 | 30 | 62 | 0.92 | 3.00 |
| CF600_30 | 55 | 66 | 0.73 | 2.24 |
| CF600_180 | 51 | 70 | 0.57 | 1.93 |

One can see that the Williamson-Hall analysis is in agreement with our assumptions and magnetic measurements. For both samples treated at 300°C there is no increase in the mean crystallite size or the strain, nor is there a significant decrease in the $H_C$ or increase in the $M_S$ (see Table 1). In the sample CF400_30, the small increase in crystallite size agrees with the small increase in $M_S$, and the decrease in the strain is consistent with the decrease in $H_C$. The results of the Williamson-Hall analysis on the samples treated at 600°C are also consistent with the magnetic measurements.

The TEM images also support our explanations for the decrease of $H_C$ and the increase of $M_S$, and are consistent with the Williamson-Hall analysis. In Figure 4 we show the TEM images of the more strained samples: CF0; CF300_30 and CF300_180. One can note contrasts in the nanoparticle images that are characteristic of strained material [11]. Many regular dislocations (see Figure 4) in the nanoparticles were also observed. These structural defects are moiré fringes [11,12] caused by the dislocation of different crystalline planes (see Figure 4E).

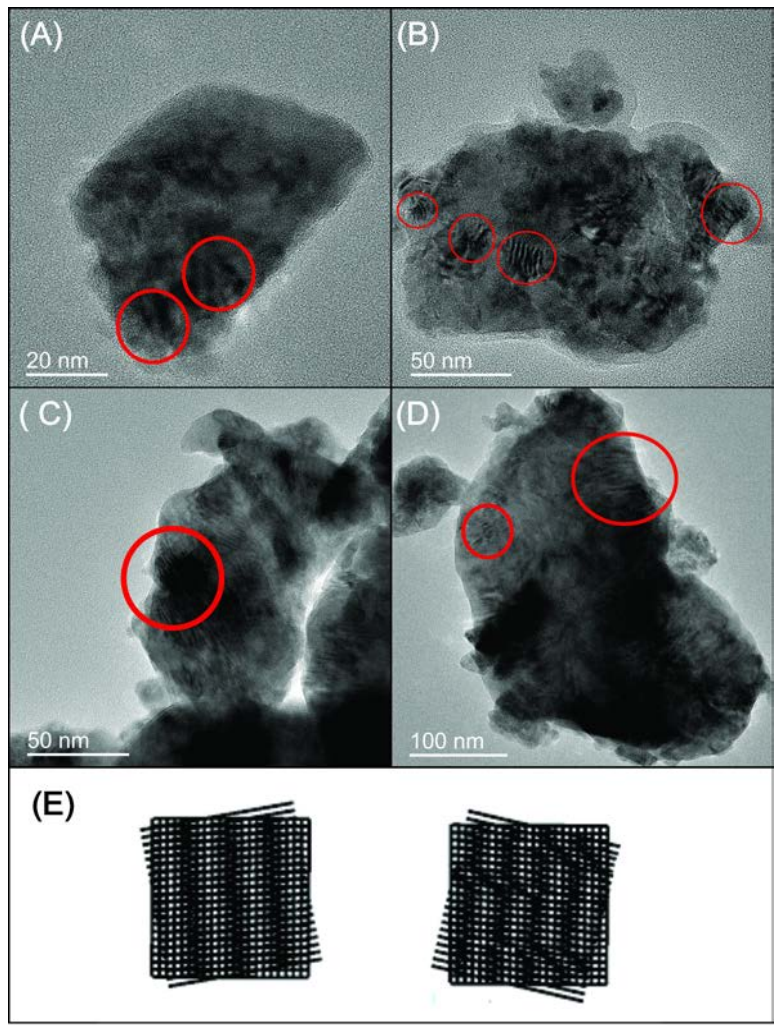

Figure 4 – TEM images of (A and B) sample CF, (C) CF300_180, (D) CF300_30, (E) regular and irregular squematic pictures of moiré fringes. The circled areas show some moiré fringes in the nanoparticles.

Some structural defects (regular dislocations) were observed in the images obtained from sample CF400_30, Figure 5(A). For images of the samples treated at 600°C (see Figure 5 B, C, and D), we did not observe any dislocation as was observed in other samples. Analysis of size distribution indicates that there is an increase in the mean size when comparing the sample CF0 with the sample CF600_180, which is compatible with our explanation.

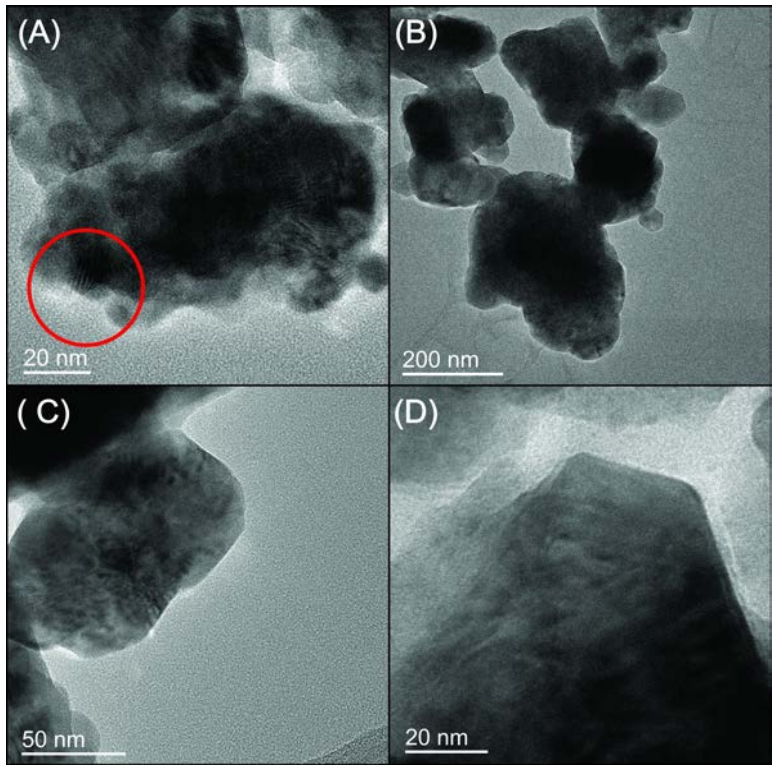

Figure 5 - TEM images of (A) sample CF400_30, (B) CF600_180, (C and D) CF600_30. The circled area shows the moiré fringes in the nanoparticle.

## Conclusion

The magnetic coercivity of the milled cobalt ferrite nanoparticles, as expected, is affected by the thermal treatment. This behaviour is associated with the decrease of microstructural strain and the density of structural defects, as confirmed by Williamson-Hall analysis and TEM measurements.

The increase in $M_S$ observed in the samples treated mainly at $600^oC$ is associated with the increase of the nanoparticle grains. This assumption is in agreement with the Williamson-Hall analysis.

To Summarize, high-coercivity cobalt ferrite induced by mechanical milling process is affected by thermal treatment. For temperatures up to $300^oC$ the coercivity is slightly affected, but the result depends on the duration of the treatment. Also, there is not a significant increase in mean size of the nanoparticles grains. Thermal treatment at temperatures equal to or higher than $600^oC$ is not recommended, even for a short duration, because a significant decrease in coercivity and the increase of the mean nanoparticles size occurs.

**Acknowledgments**

This work was supported by the following Brazilian funding agencies CAPES (project #2233/2008) and FAPEMAT (project # 850109/2009).